\documentclass[useAMS,usenatbib]{mn2e}

\def\gsim{\ \raise 3pt \hbox{$>$} \kern -8.5pt \raise -2pt \hbox{$\sim$}\ }
\def\lsim{\ \raise 3pt \hbox{$<$} \kern -8.5pt \raise -2pt \hbox{$\sim$}\ }

\usepackage{graphicx}

\begin{document}
   \title{Diffusive synchrotron radiation from extragalactic jets}

\author[G. D. Fleishman]{G. D. Fleishman$^{1,2}$\thanks{E-mail:
gregory@sun.ioffe.ru}\\
$^{1}$National Radio Astronomy Observatory, Charlottesville,
VA 22903, US\\
$^{2}$Ioffe Institute for Physics and Technology, St.Petersburg,
194021,  Russia}

\date{Accepted 2005  October 13. Received 2005 October 13; in original form 2005 September 20}

\pagerange{\pageref{firstpage}--\pageref{lastpage}} \pubyear{2005}

\maketitle

\label{firstpage}

\begin{abstract}
Flattenings of nonthermal radiation spectra observed from knots
and interknot locations of the jets of 3C273 and M87 in UV and
X-ray bands are discussed within modern models of magnetic field
generation in the relativistic jets. Specifically, we explicitly
take into account the effect of the small-scale random magnetic
field, probably present in such jets, which gives rise to emission
of Diffusive Synchrotron Radiation, whose spectrum deviates
substantially from the standard synchrotron spectrum, especially
at high frequencies. The calculated spectra agree well with the
observed ones if the energy densities contained in small-scale and
large-scale magnetic fields are comparable. The implications of
this finding for magnetic field generation, particle acceleration,
and jet composition are discussed.
\end{abstract}


\section{Introduction}

Relativistic extragalactic jets are known to provide very
efficient acceleration of relativistic electrons up to
Lorentz-factors $\gamma\sim\!10^6-10^7$ or higher
\citep{Heavens_Meisenheimer_1987}. Quasi-exponential cut-offs
found in many synchrotron sources in the infrared (IR), optical,
or ultraviolet (UV) bands
\citep{Rieke_etal_1982,Roeser_Meisenheimer_1986,Meisenheimer_Heavens_1986,Keel_1988}
are in a good agreement with the idea of a maximum  energy of
accelerated electrons, which results from the balance between the
efficiency of the acceleration mechanism and synchrotron losses.

The presence of a high-energy cut-off in the energetic spectrum of
relativistic electrons results naturally in a progressive spectral
softening as the frequency increases in the region of the
synchrotron cut-off, which indeed has been observed. However,
recent observations \citep{Jester_etal_2005} of the radio-to-UV
spectra of the jet in 3C273 performed with the highest angular
resolution achieved so far (0".3) revealed significant flattening
of the radiation spectra in the UV band from most of the jet
locations, including both knots and inter-knot regions.

This finding of an additional UV spectral component cannot be
easily accommodated within models of synchrotron emission produced
by a single population of relativistic electrons
\citep{Jester_etal_2005} and requires either a distinct secondary
component of relativistic particles, and/or a different radiative
process, dominating the UV excess. Either of these possibilities
suggests that jet models should include additional physical
processes involving particle acceleration or the radiation
mechanism. Given similar spectral behavior found in the jet of M87
in the optical to X-ray transition
\citep{Perlman_etal_2001,Marshall_etal_2002,Waters_Zepf_2005},
this problem seems to be of a general interest for jet physics.

The idea of a secondary population of the relativistic electrons
is discussed in some detail by \cite{Jester_etal_2005} who show
that it imposes rather stringent new requirements on the
acceleration mechanism involved. Here we envision an alternate
possibility predicted theoretically almost 20 years ago
\citep{Topt_Fl_1987}: that the observed spectral flattening in
certain jets is an \emph{intrinsic property} of the \emph{emission
mechanism}. Specifically, we explore the consequences of the
presence of small scale random magnetic fields for the synchrotron
radiation mechanism. Observational evidence that such fields exist
in the jet volume has been discussed by \cite{Hughes_2005} and
references therein. Moreover, recent models of magnetic field
generation in relativistic sources in general
\citep{Kazimura_1998,Medvedev_Loeb_1999,Nishikawa_etal_2003,Nishikawa_etal_2005,
Jaroshek_etal_2004,Jaroshek_etal_2005,Hededal_Nishikawa_2005}, and
in extragalactic jets in particular
\citep{Honda_Honda_2002,Honda_Honda_2004}, predict that the random
magnetic field produced is extremely small-scale, with a typical
correlation length as small as the plasma skin depth or less,
which can be less than the coherence length of synchrotron
emission:

\begin{equation}
\label{Syn_coh_length}
  l_s=\frac{mc^2}{eB}=\frac{c}{\omega_{Be}}.
\end{equation}
Here, $e$ and $m$ are the electron charge and mass, and
$\omega_{Be}={eB}/{mc}$ is the electron gyrofrequency. Generation
of electromagnetic emission by relativistic electrons moving in
small-scale magnetic fields is known to differ from the case of
emission by electrons in a uniform (large-scale) magnetic field
\citep{LL_2}.

In this paper, therefore, we address the question of whether the
presence of the small-scale random magnetic field in the jet
volume can account for the observed high-frequency excesses in the
jet spectra, and conclude that the small-scale field with a level
comparable with the  level of  large-scale field is  indeed
naturally capable of providing the observed spectral flattening.

\section{Diffusive synchrotron radiation}

The effect of the spatial scale of the magnetic field on the
corresponding electromagnetic radiation produced by fast electrons
is provided by non-local nature of the emission. Indeed, the
elementary emission pattern of the synchrotron radiation is
accumulated over a finite part of the particle trajectory of the
order of the coherence length $l_s$. Accordingly, if the magnetic
field experiences variations over this length, the particle
trajectory deviates from a circular trajectory, which results in a
radiation spectrum rather different from one formed in the uniform
or large-scale magnetic field. We will refer the synchrotron
radiative process in the presence of small-scale magnetic fields
as \emph{Diffusive Synchrotron Radiation} (DSR), since the
particle random walks due to its interaction with the random
field.

The theory of DSR was developed some time ago by
\cite{Topt_etal_1987} and \cite{Topt_Fl_1987}; see also the recent
review and developments by \cite{Fl_2005b,Fl_2005a}. When both
large-scale regular and small-scale random fields are present in a
volume, the resulting radiation spectrum is composed of two
contributions. One component results from the large-scale field
and is essentially the normal synchrotron spectrum. The second
component is DSR (called also ``3D jitter radiation'',
\cite{Hededal_2005}) resulting from the interaction of the
ultrarelativistic electrons with small-scale random fields. If the
distribution of ultrarelativistic electrons is characterized by a
maximum Lorentz factor $\gamma_m$, first component dominates for
frequencies $f<f_{Be}\gamma_m^2$, where $f_{Be} =
\omega_{Be}/2\pi$, while for frequencies $f>f_{Be} \gamma_m^2$,
the synchrotron spectrum decreases exponentially with frequency,
and the contribution of DSR becomes the dominant one.

The effect of small-scale magnetic inhomogeneities on the
radiation spectrum produced by a relativistic electron is caused
by the change in the electron trajectory compared with the regular
gyration it would experience in a large-scale magnetic field. In
the presence of a random field superimposed on the regular field,
the particle trajectory experiences random fluctuations
superimposed on the regular motion. In essence, these fluctuations
represent an incoherent superposition of oscillations with various
frequencies/wavelengths. Because of relativistic transformation,
each of these elementary oscillations with the scale $l=2\pi/k$
results in emission at the frequency $\omega \approx kc\gamma^2$.

To be more specific, let us define the statistical properties of
the small-scale random magnetic field $\mathbf{B_{st}}$ by means
of the (two-point) second-order correlation function
\begin{equation}
\label{Corr_2} K_{\alpha
\beta}^{(2)}(\mathbf{R},T,\mathbf{r},\tau) = \left<B_{st,
\alpha}(\mathbf{r}_1,t_1)B_{st,\beta}(\mathbf{r}_2,t_2)\right>,
\end{equation}
where $\mathbf{R} =(\mathbf{r}_1 + \mathbf{r}_2)/2$, $\mathbf{r}
=\mathbf{r}_2 - \mathbf{r}_1$, $T =(t_1 + t_2)/2$, and $\tau =t_2
- t_1$. Assume that the random field is statistically uniform in
space and time, and take the Fourier transform of the correlator
$K_{\alpha \beta}^{(2)}(\mathbf{r},\tau)$ over spatial and
temporal variables $\mathbf{r}$ and $\tau$, which yields the
spectral representation of the random field
\begin{equation}
\label{Corr_2_spectr} K_{\alpha \beta}(\mathbf{k}, \omega) = \int
{d\mathbf{r} d \tau \over (2 \pi)^4} e^{i(\omega \tau -
\mathbf{kr})} K_{\alpha \beta}^{(2)}(\mathbf{r},\tau).
\end{equation}
For the isotropic wave turbulence we  find  easily
\citep{Fl_2005b}
\begin{equation}
\label{Corr_2_isotr} K_{\alpha \beta}(\mathbf{k},
\omega)=\frac{1}{2}K(\mathbf{k}) \delta(\omega -
\omega(\mathbf{k}))\left(\delta_{\alpha \beta} - {k_{\alpha}
k_{\beta} \over k^2}\right),
\end{equation}
where $K(\mathbf{k})$ describes the spatial spectrum of the
small-scale random magnetic field. The correlator
(\ref{Corr_2_isotr}) satisfies Maxwell's equation $\nabla \cdot
\mathbf{B}_{st} = 0$, since the tensor structure of the correlator
is orthogonal to the $\bf k$ vector: $k_{\alpha}K_{\alpha
\beta}=0$.

For the following modelling  we  adopt a power-law spectrum of the
random magnetic field \citep[see,
e.g.,][]{turbulence,Hededal_2005}:
\begin{equation}
\label{power_spectr}
 K(\mathbf{k}) = \frac{A_{\nu}}{k^{\nu+2}},\ A_{\nu}= {(\nu-1)  k_{min}^{\nu-1}
 \left<B_{st}^2\right> \over 4 \pi}, \ k_{min}< k <
k_{max},
\end{equation}
where $\nu$  is the spectral index of the turbulence, and the
spectrum  $K(\mathbf{k})$ is normalized to $d^3k$:
\begin{equation}
\label{spectr_norm} \int_{k_{min}}^{k_{max}} K(\mathbf{k}) d^3k =
\left<B_{st}^2\right>, \ при \ k_{min} \ll k_{max}, \ \nu >1,
\end{equation}
where $\left<B_{st}^2\right>$ is the mean square of the
small-scale random field, $k_{min} \sim {eB_{ls}}/{mc^2}$,
$B_{ls}$ is the characteristic value of the \emph{regular
large-scale} magnetic field at the source. Therefore, the
properties of the isotropic random magnetic field are described
here by three main measures: rms strength of the field $B_{st}$,
the main scale $L=2\pi/k_{min}$, and the spectral index $\nu$,
which characterizes the distribution of the magnetic energy over
spatial scales.

The radiation spectrum produced by a single relativistic particle
with the Lorentz-factor $\gamma$ in the presence of a regular
field and the adopted small-scale random field is described by the
asymptote \citep{Topt_Fl_1987,Fl_2005b}
\begin{equation}
\label{I_case2_full_2}
 I_{\omega}  = \frac{2^{\nu}(\nu-1)(\nu^2 +7\nu +8)}
 {\nu(\nu+2)^2(\nu+3)}\ \frac{e^2}{c} \
 \frac{\omega_{st}^2\omega_{0}^{\nu-1}\gamma^{2\nu}}
 {\omega^{\nu}}
\end{equation}
at the high-frequency range $\omega \gg \omega_{Be} \gamma^2$,
where $\omega_{st}^2=e^2\left<B_{st}^2\right>/(mc)^2$,
$\omega_{0}=k_{min}c$. Evidently, the intensity of radiation in
the range between $\omega$ and $\omega+d\omega$ is specified by
the energy density of the random field  in the range between a
corresponding $k$ and $k+d k$. This eventually results in the
spectrum $I_{\omega} \propto \omega^{-\nu}$ (\ref{I_case2_full_2})
with the spectral index equal to the spectral index of the random
field $\nu$ in (\ref{power_spectr}) \citep{Topt_Fl_1987,Fl_2005b}
at the frequencies larger than $\omega_{Be}\gamma^2$, where the
standard synchrotron radiation decreases exponentially.
Accordingly, the small-scale spatial harmonics of the random field
with $k > \omega_{Be}/c$ are responsible for this spectral range,
$\omega > \omega_{Be}\gamma^2$, while larger scale harmonics ($k <
\omega_{Be}/c$) are less important, since the spectral range
$\omega < \omega_{Be}\gamma^2$ is specified by standard
synchrotron radiation in the regular large-scale field.

\begin{figure*}
\includegraphics[height=2.9in]{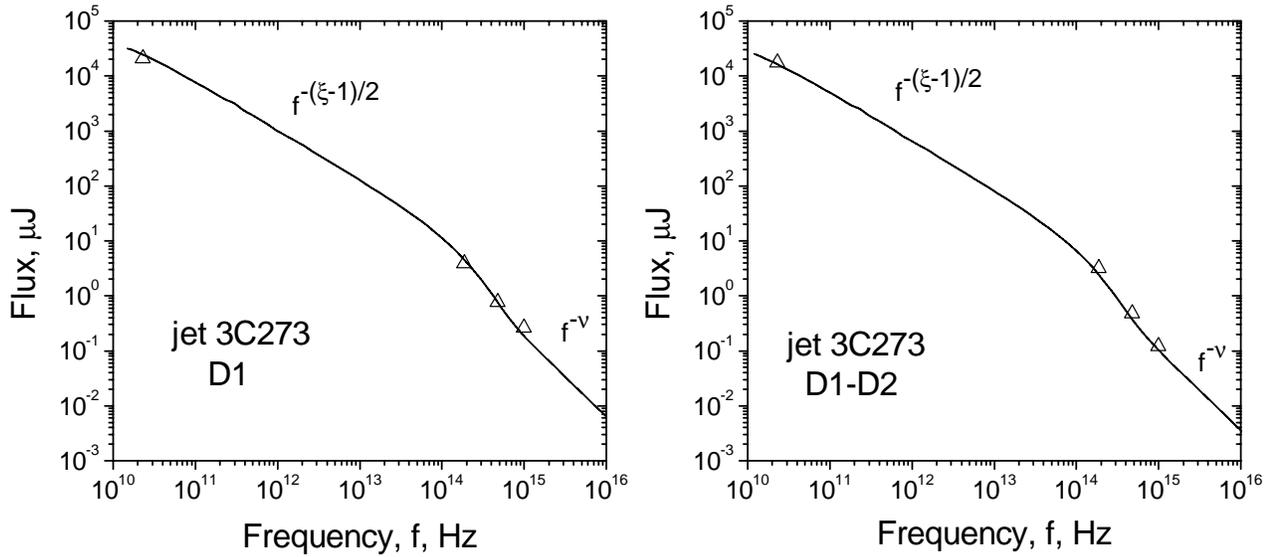}
\caption{Radio to UV spectra  (triangles) of knot D1 (left) and
inter-knot D1-D2 region (right) of the jet of 3C273 from Jester et
al. (2005)  and model DSR spectra (solid curves) for $B_{st}=1.2
B_{ls}$ and $\nu=1.5$. Parameters specifying radio to optical
synchrotron part of the spectrum are the same as in Jester et al.
(2005). }
 \label{spectra_3C273}
\end{figure*}

The entire radiation spectrum produced by an \emph{ensemble} of
relativistic electrons in the presence of small-scale and
large-scale magnetic fields is given by integration of the
single-particle spectrum (described by formulae (35) of
\citep{Topt_Fl_1987}, see also Eq. (30) in \cite{Fl_2005b}) over
the distribution of relativistic electrons over energy. We adopt a
simple power-law distribution of the relativistic electrons over
Lorentz-factor $\gamma$ with a sharp cut-off at $\gamma_m$:
\begin{equation}
\label{distr_fun_power}
  dN_e(\gamma) = (\xi-1)N_e \gamma_1^{\xi-1} \gamma^{-\xi}, \ \ \gamma_1 \le \gamma
  \le \gamma_m,
\end{equation}
where $N_e$  is the number density of relativistic electrons with
energies ${\cal E} \ge mc^2 \gamma_1$, $\xi$ is the power-law
index of the distribution.

Actually, the full electron spectrum in the extragalactic jets is
not a simple power-law, which is evident from the turn-over of the
radiation spectrum in the microwave range. However, those
turn-overs can be easily  interpreted in terms of synchrotron
losses \citep[][and references therein]{Jester_etal_2005} and are
not discussed here. Therefore, we select a large enough value of
$\gamma_1$, corresponding to the frequencies \emph{above} the
microwave turn-over, where the model of the simple power-law
distribution (\ref{distr_fun_power}) looks appropriate.
Accordingly, the index $\xi$ corresponds to high-energy indices as
defined in \citep{Jester_etal_2005}.

For ensemble of radiating electrons with a sharp cut-off
(\ref{distr_fun_power}), the DSR contribution related to the
small-scale random field will be the dominant one at the
frequencies $\omega > \omega_{Be}\gamma_m^2$, where the normal
synchrotron radiation from the whole \emph{ensemble} decreases
exponentially. The corresponding DSR asymptote (which may be
evaluated by the integration of single particle asymptote
(\ref{I_case2_full_2}) with the electron distribution over the
Lorenz-factors (\ref{distr_fun_power})) has the form:
\begin{equation}
\label{P_high_random_ans}
 $$ P_{\omega}  = \frac{2^{\nu}(\xi-1)(\nu-1)(\nu^2 +7\nu +8)}
 {\nu(2\nu-\xi+1)(\nu+2)^2(\nu+3)}\ \frac{e^2N_e \gamma_1^{\xi-1}}{c} \
 \frac{\omega_{st}^2\omega_{0}^{\nu-1}\gamma_m^{2\nu-\xi+1}}{\omega^{\nu}}.
\end{equation}

Although the small-scale random magnetic field is defined above by
three free parameters, two of them can be substantially
constrained prior to the spectrum modelling. Since only
small-scale spatial harmonics ($k>\omega_{Be}/c$) are important at
frequencies  $\omega > \omega_{Be}\gamma_m^2$, we can adopt
$k_{min}  = \omega_{Be}/c \equiv eB_{ls}/(mc^2)$ or
$\omega_{0}=\omega_{Be}$. Then, we adopt $\nu=1.5$ in agreement
with available models of magnetic turbulence
\citep{turbulence,Hededal_2005} as well as with X-ray spectral
slope for a few locations of the  M87 jet
\citep{Marshall_etal_2002}. Therefore, only one free parameter,
the small-scale magnetic field rms strength $B_{st}$, is left for
the purpose of the spectrum fitting.

Fig.\ref{spectra_3C273} displays full DSR spectra (numerically
calculated with general formulae obtained by \cite{Topt_Fl_1987})
that agree well with the optical-UV observations of one knot and
one inter-knot location in the jet of 3C273. The rms value of the
small-scale random magnetic field is adopted to be $B_{st}=1.2
B_{ls}$, where $B_{ls}$ is the large-scale magnetic field, for
both panels. The assumed presence of the small-scale magnetic
field is, therefore, responsible for the observed flattening in
the optical-UV transition, while the parameters defining
lower-frequency part of the spectrum (i.e., the critical frequency
of the synchrotron radiation, $f_c=1.5f_{Be}\gamma_m^2$ \citep[as
defined in][]{Kellerman_1964}, and the electron spectral index
$\xi$) are essentially similar to those determined in
\citep{Jester_etal_2005}, since the small-scale random field has
only a weak effect on the spectrum at $f<f_c$ if $B_{st} <
3B_{ls}$ \citep{Fl_2005b}.

\begin{figure*}
\includegraphics[height=2.9in]{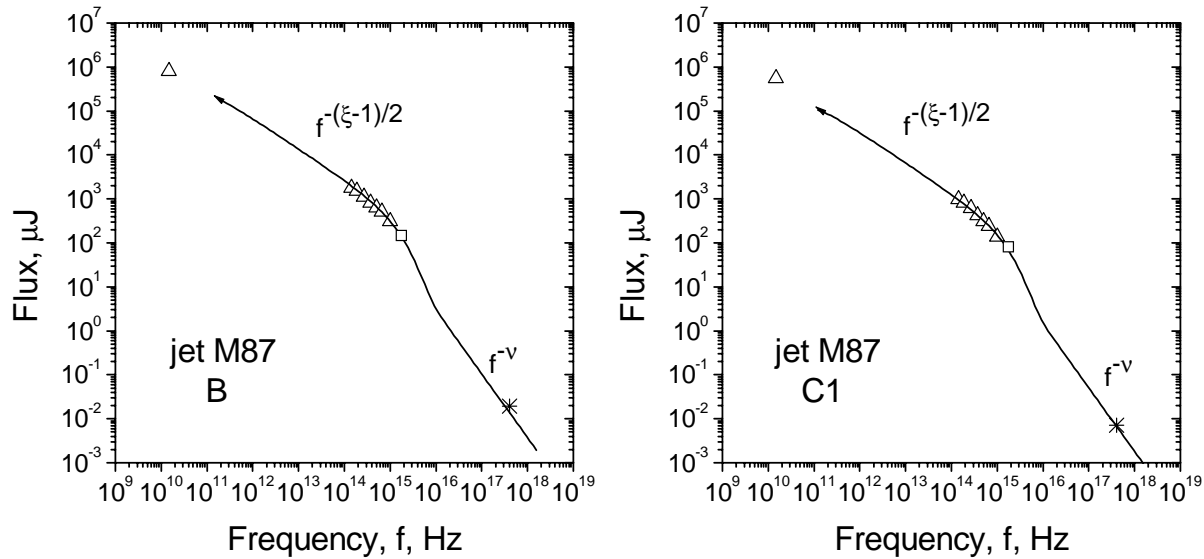} \caption{Radio to X-ray spectra (symbols)
of knots B (left) and  C1  (right) of the jet of M87 and model DSR
spectra (solid curves) for $B_{st}= B_{ls}$ and $\nu=1.5$. The
radio to UV data (triangles) are taken from Perlman et al. (2001),
UV data at $2\times 10^{15}$ Hz (squares) are from Waters \& Zepf
(2005), while X-ray data (asterisks) are from Marshall et al.
(2002).} \label{spectra_M87}
\end{figure*}

Similarly, Fig.\ref{spectra_M87}  displays DSR spectra
superimposed on the observational data for two locations of the
jet of M87
\citep{Perlman_etal_2001,Marshall_etal_2002,Waters_Zepf_2005}.
Remarkably, the X-ray observations are apparently consistent with
DSR calculated for the parameters similar to the case of the jet
of 3C273 (the only minor difference is $B_{st}= B_{ls}$ in place
of $B_{st}=1.2 B_{ls}$). It is important to note that the X-ray
spectral index determined by \cite{Marshall_etal_2002} for a few
spatially resolved locations, $\alpha_X \simeq 1.5$, is in
excellent agreement with the standard models and observations of
the turbulence
spectra 
\citep{turbulence,Laz_Ber_2005,Schek_Cowley_2005}. We emphasize
that the DSR spectrum gives equally good fits to other knots
through the jets (where X-ray data is available) with the same
ratio of the small-scale to large-scale field, while with varying
values of the synchrotron critical frequency $f_c$ or/and electron
spectral index.

\section{Discussion and Conclusions}

\cite{Marshall_etal_2002}, \cite{Jester_etal_2005}, and
\cite{Waters_Zepf_2005} compared the existing models of jet
emission with the broadband spatially resolved observations of the
jets of M87 and 3C273, and ruled out the one-component synchrotron
models as well as inverse Compton models. They identify and
clearly describe serious shortcomings of current jet models, which
fail to provide a reasonable fit to the observations. In
particular, \cite{Marshall_etal_2002} emphasize that the X-ray
spectral slope, $\alpha_X \simeq 1.5$, typical for a few locations
through the  M87 jet, is inconsistent with the available models,
while \cite{Perlman_Wilson_2005} propose  the filling factor of
the emitting material to change with frequency and location along
the jet to account for the observed spectral energy distribution.
In contrast, the DSR spectrum produced by a single power-law
electron population in the uniform (on average) source provides
excellent fits to the data in all cases considered.

The presented interpretation of the flattening of the non-thermal
radiation spectra observed from spatially resolved locations in
two extragalactic jets, 3C273 and M87, is self-consistent. Indeed,
contemporary models of the magnetic field structure suggest that
the presence of relatively strong small-scale random magnetic
fields in the jet volume is likely. The most recent
particle-in-cell simulations
\citep{Nishikawa_etal_2003,Nishikawa_etal_2005,
Jaroshek_etal_2004,Jaroshek_etal_2005,Hededal_Nishikawa_2005,Hededal_2005}
clearly indicate that strong small-scale inhomogeneities are
efficiently generated at the shock front and exist behind the
front, although the currently available numerical capacities are
still insufficient to reliably track the evolution of this
turbulence far away from the shock front. Nevertheless, the
presence of highly enhanced level of small-scale magnetic
inhomogeneities is needed for efficient pitch-angle scattering of
relativistic electrons required in many models of particle
acceleration. However, this elusive but important quantity has
never been reliably estimated for the jet volume either
theoretically or observationally.

The use of the well-established DSR theory \citep{Topt_Fl_1987}
offers a straightforward way of estimating properties of the
small-scale magnetic inhomogeneities. The synchrotron radiation
spectrum produced by ultrarelativistic electrons in the jet is
modified by the presence of small scale random magnetic fields. In
particular, taking explicit account of the perturbations to the
electron trajectories as a result of small scale magnetic fields,
here referred to as DSR, yields a high-frequency power-law
component to the spectrum that dominates over regular synchrotron
emission above the exponential cutoff.

The proposed DSR model with only one more free parameter compared
with standard synchrotron models agrees excellently with
observations of these two jets if $B_{st} \sim B_{ls}$, which
assumes a highly turbulent magnetic field in the jet volume.
Curiously, \cite{Perlman_Wilson_2005} recently reported that
spatial peaks of the X-ray flux coincide with minima of optical
polarization in the M87 jet. Therefore, the presence of a tangled
magnetic field (even though not necessarily with such small scales
as required to produce DSR) is a favorable condition to produce
X-ray emission from the jet volume, which is in qualitative
agreement with the DSR model. We emphasize that the small-scale
field alone would not produce the exponential region of the
radiation spectrum \citep{Fl_2005b}, so the presence of the
large-scale magnetic field is necessary.

Nevertheless, it would be highly desirable to have any additional
confirmation in favor of the proposed interpretation besides the
spectral fit itself, e.g., obtained from the polarization
measurement. Evidently, the DSR generated in the presence of
isotropic magnetic fluctuations adopted here is unpolarized
emission. However, to be speculative, one may suppose some
anisotropic distributions of the magnetic turbulence
\citep{Laz_Ber_2005} given that the jets themselves are highly
anisotropic objects. The polarization patterns (the position
angle, the degree of polarization, and its spectral behavior) will
be eventually set up by the type and strength of the anisotropy of
the random magnetic field \citep{Topt_Fl_1987b} and will be
generally different from both standard synchrotron emission and
inverse Compton emission, which may therefore allow a clear
distinction to be made between mechanisms. One of the simplest
examples of the anisotropic random field is the turbulence
composed of random waves with 1d-distribution of ${\bf k}$-vectors
along the regular magnetic field. In this specific case the
direction of the random field will be mainly orthogonal to the
uniform magnetic field, which will result in the rotation of the
position angle by $90^o$ when the transition from standard to
diffusive synchrotron emission occurs as frequency increases
\citep{Topt_Fl_1987b}. Although the polarization measurement
needed to provide a direct test for the spectra considered in this
paper is currently unavailable in UV and X-ray bands, one may try
to study the polarization of the optical emission from those jets
and/or jet locations that display lower value of the synchrotron
high-frequency cut-off, e.g., in the IR band.

Overall, analysis of the multi-wavelength spectra from several
locations along two extragalactic jets (3C273 and M87)
demonstrates that the spectra considered are well described by
DSR, assuming $B_{st} \sim B_{ls}$. This property may result from
a saturated state of the microscopic process responsible for the
magnetic field generation in the jets, e.g., from the nonlinear
stage \citep{Jaroshek_etal_2005} of the Weibel instability
\citep{Weibel_1959}. In principle, this can provide us with some
constraints on the jet composition. In particular, the model of
\cite{Honda_Honda_2002,Honda_Honda_2004} links the characteristic
width of the current filaments (and, accordingly, the correlation
scale of the magnetic field produced) with the rate of
electron-positron pair production. Therefore, the ratio of the
small-scale to large-scale field will differ for various regimes
of the pair generation, which makes the observational measurements
of this ratio of primary importance. Moreover, a precise
measurement of the spectral indices of the radiation in the
high-frequency range gives (potentially)  direct information on
the spectrum of short-wave turbulence in relativistic sources,
which is highly relevant to the models of magnetic field
generation and relativistic particle acceleration. We conclude
that DSR from small scale magnetic fields, which are a natural
outcome of current jet models, offers a plausible explanation for
the observed spectral flattening in the emission from certain
extragalactic jets, one which does not appeal to the presence of a
secondary population of accelerated particles.

\section*{Acknowledgments}

The National Radio Astronomy Observatory is a facility of the
National Science Foundation operated under cooperative agreement
by Associated Universities, Inc. This work was supported in part
by  the Russian Foundation for Basic Research, grants
No.03-02-17218, 04-02-39029. I am strongly grateful to T.S.
Bastian for his numerous comments for the paper, to P.A.Hughes for
a number of important suggestions, and to A. Bridle, B. Cotton,
and K. Kellerman for discussion of the paper.



\label{lastpage}

\end{document}